\documentclass[onecolumn,showpacs,preprintnumbers,amsmath,amssymb]{revtex4}

\usepackage{amsmath}    
\usepackage{graphicx}   
\usepackage{graphics}
\usepackage{dcolumn}
\usepackage{bm}

\usepackage{multirow}

\usepackage{latexsym}
\usepackage{amssymb}
\usepackage{amsmath}

\usepackage{color}
\usepackage{epstopdf}

\usepackage[normalem]{ulem}

\begin{document}
\title{Antiferromagnetic spin cantings as a driving force of ferroelectricity in multiferroic Cu$_2$OSeO$_3$}
\author{Viacheslav A. Chizhikov\footnote{email: chizhikov@crys.ras.ru} and Vladimir E. Dmitrienko\footnote{email: dmitrien@crys.ras.ru}}
\affiliation{FSRC ``Crystallography and Photonics'' RAS, 119333 Moscow, Russia}


\begin{abstract}
Ferroelectric properties of cubic chiral magnet Cu$_2$OSeO$_3$ can emerge due to the spin noncollinearity induced by antiferromagnetic cantings. These cantings are the result of the Dzyaloshinskii--Moriya interaction and in many ways similar to the ferromagnetic cantings in weak ferromagnets. An expression for the local electric polarization is derived, including terms with gradients of magnetization $\mathbf{M}(\mathbf{r})$. When averaged over the crystal the electric polarization has a non-vanishing part associated with the anisotropy of the crystal point group 23. In the framework of the microscopic theory, it is shown that both scalar and vector products of spins, $(\mathbf{s}_1\cdot\mathbf{s}_2)$ and $[\mathbf{s}_1\times\mathbf{s}_2]$, can give contributions of the same order in the electric polarization.
\end{abstract}
\maketitle


\section{Introduction}
\label{sec:intro}

The development of new areas of electronics, such as spintronics, photonics and quantum computing, requires the discovery of new advanced materials and the investigation of their unusual physical properties and combinations thereof. Such combination provides the simultaneous presence of ferromagnetic, ferroelectric and/or ferroelastic properties. It is of great interest, when these properties do not co-exist independently but arise at a time as a result of a spontaneous symmetry breaking, usually at low temperatures. Such crystals are called type-II multiferroics, and for them in contrast to the type-I multiferroic it is possible to manage one property, directly affecting another one, e.g., to change the magnetic structure using an electric field \cite{Pyatakov2012}.

\begin{figure}[h]
\includegraphics[width=7cm]{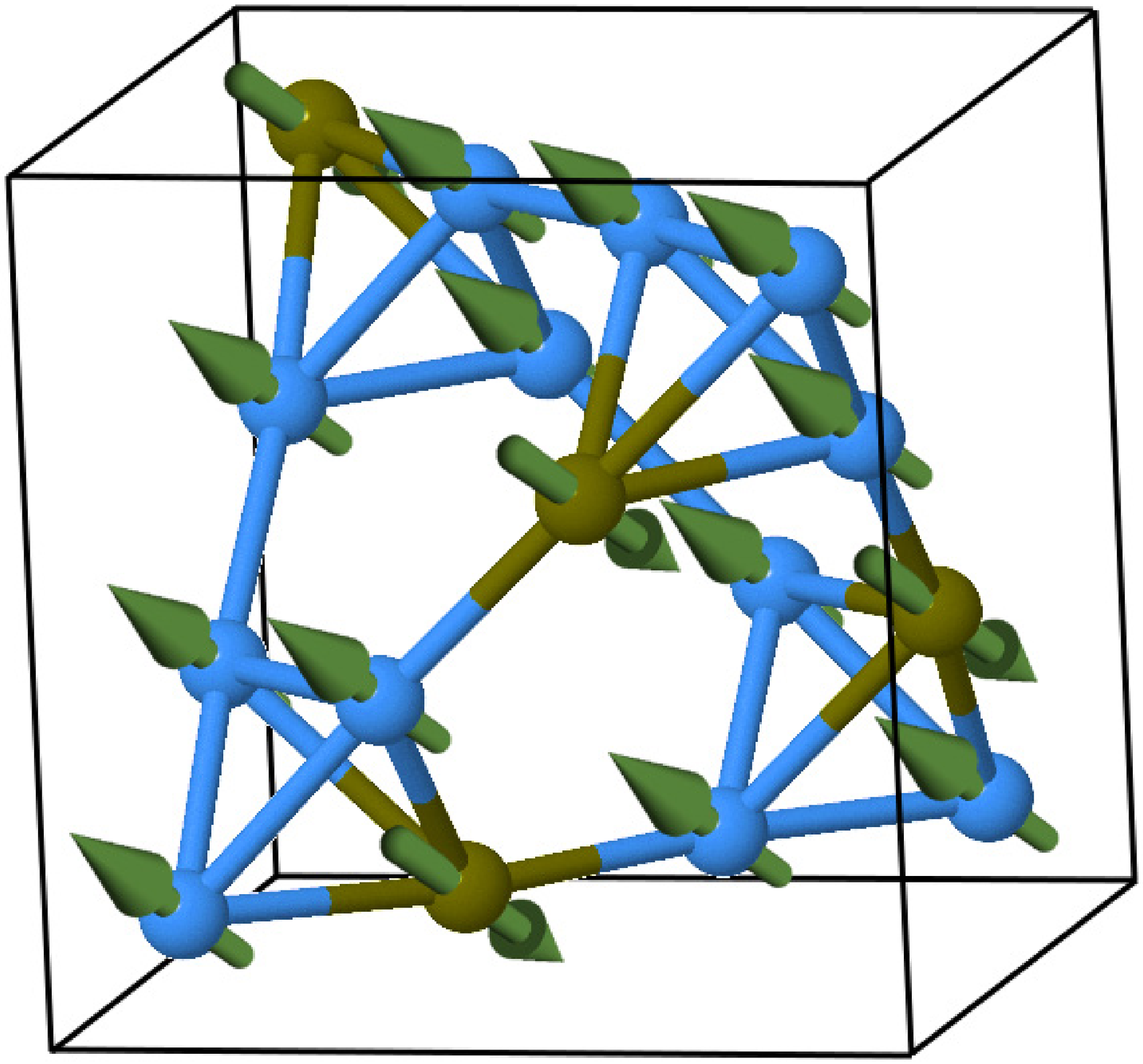}
\caption{\label{fig:cu2oseo3} (Color online) The copper sublattice of Cu$_2$OSeO$_3$. 16 atoms in the unit cell are distributed between two non-equivalent positions: 4$a$ Cu-I (brown) and 12$b$ Cu-II (blue). In the magnetically ordered state, the spins in positions Cu-I and Cu-II have the opposite directions, making the crystal a ferrimagnet. The total magnetization is directed along the spins of Cu-II.}
\end{figure}

Among these materials the cubic crystal Cu$_2$OSeO$_3$ holds a special place. The crystal possesses non-symmorphic cubic space group $P2_13$. Its unit cell contains 16 magnetic copper atoms, distributed between two non-equivalent atomic positions (Fig.~\ref{fig:cu2oseo3}). Due to the absence of an inversion center the magnetic structure of the crystal is rather unusual. Recent studies have shown that in addition to the trivial helical structure, similar to the helicoid in cholesteric liquid crystals, in Cu$_2$OSeO$_3$ a double twisted phase can occur \cite{SekiSci2012,AdamsPRL2012,SekiPRB2012,OnosePRL2012,SekiPRB2012-2,Belesi2012,Janson2014,Mochizuki2015,OvchinnikovUFN2014}, which is a triangular lattice of 2D skyrmions, resembling a network of the Abrikosov vortices in the type-II superconductors. Earlier the A-phase has already been observed in the itinerant magnets of structural type $B20$, such as MnSi \cite{OvchinnikovUFN2014,Grigoriev2006a,Muhlbauer2009,Munzer2010,Adams2011}, but unlike the latter Cu$_2$OSeO$_3$ is an insulator and therefore may also be ferroelectric. Knowing the average direction of polarization vector can be used to provide control of the magnetic structure, in particular to stabilize or destruct the A-phase by an electric field \cite{Mochizuki2015,White2012}.

The appearance of an electric polarization in the type-II multiferroics is often associated with the spin noncollinearity. In the chiral magnets, such as Cu$_2$OSeO$_3$, the cause of the noncollinearity is a global twist of the magnetic structure. By applying a magnetic field, it is possible to remove the twist and, consequently, the spin noncollinearity induced by it. It would seem that in this case the electric polarization should disappear, but an experimental data prove the contrary. Indeed,  there remains a polarization, which depends on the orientation of the magnetization with respect to the crystallographic axes \cite{SekiPRB2012-2}. In order to explain this effect, a single-spin mechanism has been developed of the origin of ferroelectricity, where the local polarization corresponding to each individual atom is determined by the orientation of its magnetic moment relative to the cubic crystal axes \cite{SekiPRB2012-2,Mochizuki2015,Ruff2015}. However, as shown in \cite{Chizhikov2015}, there is another source of noncollinearity, which is due to the antiferromagnetic spin cantings caused by the Dzyaloshinskii--Moriya interaction. This noncollinearity does not disappear even in the magnetic fields much larger then  $H_{c2}$, required for  full unwinding of the magnetic structure. As will be shown in the paper, the noncollinearity induced by the cantings can lead to observable effects similar to those predicted by the single-spin model. In addition, new terms are predicted in the expression for the electric polarization, containing magnetization gradients.

Sec.~\ref{sec:epol} shows how the spin noncollinearity may give rise to a ferroelectricity in the type-II multiferroics. Sec.~\ref{sec:2approa} provides a brief description of the magnetic properties of the chiral cubic magnets in two complementary models: the continuous phenomenological approximation and the discrete microscopic approach. In Sec.~\ref{sec:epolcub} an expression is derived for the local electric polarization induced by a magnetization field. Sec.~\ref{sec:averagepol} discusses the polarization averaged over the crystal for different magnetic structures, including the A-phase. In Sec.~\ref{sec:phenom} a phenomenological theory is suggested, which describes adequately the magnetoelectric properties of the Cu$_2$OSeO$_3$ crystal. Sec.~\ref{sec:discussion} briefly discusses the methods developed by other authors, and draw a comparison with the results obtained in the article.

\section{Electric polarization in the type-II multiferroics}
\label{sec:epol}

Unlike the type-I multiferroics, the ferroelectric properties of which are determined by a crystal structure symmetry, in the type-II multiferroics the electric polarization appears simultaneously with a magnetic ordering, i.e. at a phase transition from the high-symmetry paramagnetic state to a low-symmetry ferromagnetic one. The additional condition of ferroelectricity is the presence of magnetic ions with different charges, or, as in our case, a noncollinear magnetic structure. In the latter case, the spin noncollinearity is often associated with  gradients of the magnetization field, although, as mentioned above, the noncollinearity due to cantings (Fig.~\ref{fig:canting}a) can cause a polarization even in the homogeneous state, where spirals are untwisted by an external magnetic field.

Within the framework of the microscopic description, a local electric polarization can be associated with pairs of neighboring magnetic atoms, and different models define it either through the scalar product of two spins ($\mathbf{s}_1\cdot\mathbf{s}_2$) \cite{ChaponPRL2006,JiaPRB2007,Zobkalo2014}, or through their vector product [$\mathbf{s}_1\times\mathbf{s}_2$] \cite{ChaponPRL2006,JiaPRB2007,Zobkalo2014,KatsuraPRL2005,Arima2007,Radaelli2008}. We will keep both of these products, because further arguments show that the contributions of these terms can be approximately of the same order. Let us pick out from Eq.~(\ref{eq:Emicro}) the energy associated with the interaction of spins $\mathbf{s}_1$ and $\mathbf{s}_2$:
\begin{equation}
\label{eq:s1s2_energy}
E_{12} = -J_{12} (\mathbf{s}_1 \cdot \mathbf{s}_2) + \mathbf{D}_{12} \cdot [\mathbf{s}_1 \times \mathbf{s}_2] .
\end{equation}

Let $q$ be a charge participating in the exchange interaction of the spins. It could be a charge of one of the interacting magnetic ions, 1 or 2, or a non-magnetic ion charge, through which superexchange interaction is performed. A displacement of the charge $q$ from its equilibrium position at a distance of $\mathbf{u}$ should change the values of interaction parameters of the bond 12,
\begin{equation}
\label{J12uD12u}
\begin{array}{l}
J_{12}(\mathbf{u}) = J_{12}(0) +(\partial J_{12} / \partial u_{\alpha}) u_{\alpha}, \\
\mathbf{D}_{12}(\mathbf{u}) = \mathbf{D}_{12}(0) + (\partial \mathbf{D}_{12} / \partial u_{\alpha}) u_{\alpha},
\end{array}
\end{equation}
resulting in a change in energy (\ref{eq:Emicro}). This gives rise to a local electric dipole moment $\mathbf{p}=q\mathbf{u}$. The value of $\mathbf{u}$ can be found by minimizing the energy
\begin{equation}
\label{eq:u_energy}
E(\mathbf{u}) = -\frac{\partial J_{12}}{\partial u_{\alpha}} u_{\alpha} (\mathbf{s}_1 \cdot \mathbf{s}_2) + \frac{\partial D_{12,\beta}}{\partial u_{\alpha}} u_{\alpha} [\mathbf{s}_1 \times \mathbf{s}_2]_{\beta} +\frac12 A_{\alpha\beta} u_{\alpha} u_{\beta} ,
\end{equation}
with $A$ being a symmetric positive definite matrix, which defines the elastic energy of the displacement of charge $q$ from its equilibrium position.

The minimization of Eq.~(\ref{eq:u_energy}) gives
\begin{equation}
\label{eq:u_alpha}
u_\alpha = \tilde{v}_\alpha (\mathbf{s}_1 \cdot \mathbf{s}_2) + \tilde{T}_{\alpha\beta} [\mathbf{s}_1 \times \mathbf{s}_2]_{\beta} ,
\end{equation}
where $\tilde{\mathbf{v}}$ depends on derivatives $\partial J_{12} / \partial \mathbf{u}$, and $\tilde{T}$ contains combinations of $\partial \mathbf{D}_{12} / \partial \mathbf{u}$,
\begin{equation}
\label{eq:vT}
\tilde{v}_\alpha = (A^{-1})_{\alpha\gamma} \frac{\partial J_{12}}{\partial u_\gamma} , \phantom{xx} \tilde{T}_{\alpha\beta} = - (A^{-1})_{\alpha\gamma} \frac{\partial D_{12,\beta}}{\partial u_\gamma} .
\end{equation}
Assuming that parameters $J_{12}(\mathbf{u})$ and $\mathbf{D}_{12}(\mathbf{u})$ change significantly on the same length scale, the components of tensor $\tilde{T}$ are less than the coordinates of vector $\tilde{\mathbf{v}}$ by a factor of order of $D/J \ll 1$.

The total polarization of the bond is calculated as the sum over all charges involved in the magnetic interaction of two spins. A correction of spins $\mathbf{s}_1$ and $\mathbf{s}_2$ due to the change of the exchange parameters can also affect the energy. In order to avoid the difficulties of direct calculations and to take into account the symmetry of the crystal structure, it is convenient to introduce a phenomenological expression for the electric polarization of the bond,
\begin{equation}
\label{eq:pbond}
p_{12,\alpha} = v_\alpha (\mathbf{s}_1 \cdot \mathbf{s}_2) + T_{\alpha\beta} [\mathbf{s}_1 \times \mathbf{s}_2]_{\beta} .
\end{equation}
Here, vector $\mathbf{v}$ possesses the point symmetry of bond 12, and pseudotensor $T$ is transformed as follows: it is symmetric with respect to all the symmetry transformations, which do not permutate spins $\mathbf{s}_1$ and $\mathbf{s}_2$, and antisymmetric if the symmetry operation rearranges the spins (such a pseudotensor can be of form $T_{\alpha\beta}=T_{\alpha\beta\gamma}(\mathbf{r}_{12})_\gamma$, with $\mathbf{r}_{12}$ being the distance between atoms 1 and 2). The total electric moment density is obtained by summing the electric polarizations of all the bonds in the unit cell. Obviously, this takes into account the symmetry of the crystal. Note that, as in the case of $\tilde{T}$ and $\tilde{\mathbf{v}}$, the components of $T$ are less than the components of $\mathbf{v}$ by a factor of small parameter $D/J$. Moreover, if we consider weak, also of order of $D/J$, noncollinearity of spins, the first term in Eq.~(\ref{eq:pbond}) is greater than the second one by approximately two orders of magnitude. However, upon averaging over all the bonds within the unit cell, e.g., in the case of cubic symmetry, $\left<\mathbf{v}\right>=0$ and the contributions of the first and second terms in Eq.~(\ref{eq:pbond}) can be of the same order.

\section{Two approaches to the description of cubic helimagnets}
\label{sec:2approa}

Before turning to the ferroelectric properties of the chiral cubic magnets and their relation to the magnetic ones, a brief introduction will be done about magnetic structure of the materials and theoretical approaches describing it. As is usually the case in dealing with the problems of the condensed matter physics, all descriptive methods are divided into two main classes: (1) a macroscopic continuous approximation and (2) a microscopic theory taking into account the real structure of crystals. Consider both approaches and their consequences.

\subsection{Continuous approximation}

\begin{figure}[h]
\includegraphics[width=7cm]{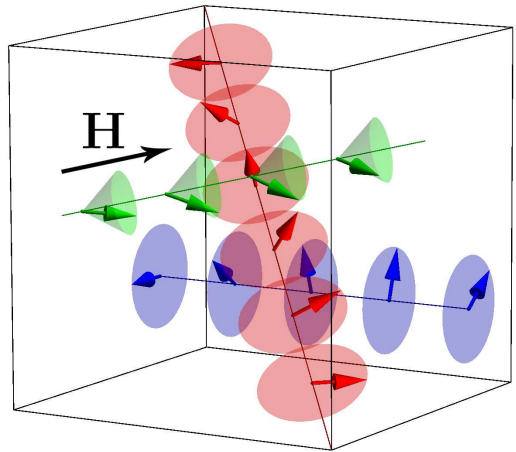}
\caption{\label{fig:helices} (Color online) In the absence of an external magnetic field, due to the cubic anisotropy there are two possible orientations of the magnetic helix, either $\left<100\right>$ (blue) or $\left<111\right>$ (red) depending on cubic anisotropy \cite{Bak80}. In a magnetic field the helix becomes conical and aligned along the field (green).}
\end{figure}

The global magnetic structure of a cubic helimagnet can be described using the phenomenological Landau---Lifshitz theory, which in the simplest case is given by the energy \cite{Bak80}
\begin{equation}
\label{eq:Ephem}
E = J \frac{\partial M_\alpha}{\partial r_\beta} \frac{\partial M_\alpha}{\partial r_\beta} + D \mathbf{M} \cdot [\boldsymbol{\nabla} \times \mathbf{M}] -  \mathbf{H} \cdot \mathbf{M} ,
\end{equation}
where $J$ is an isotropic exchange parameter, $D$ is a constant of the Dzyaloshinskii--Moriya interaction, $\mathbf{H}$ is an external magnetic field, and magnetization $\mathbf{M}$ has a constant absolute value $M_0$. Minimization of (\ref{eq:Ephem}) in the absence of a field ($\mathbf{H}=0$) gives the solution in the form of magnetic helix with the wavenumber $k=D/2J$, and the magnetization vector $\mathbf{M}$ rotating in the plane perpendicular to the helix axis $\mathbf{n}$. The direction $\mathbf{n}$ is not fixed by the energy (\ref{eq:Ephem}), but it can be easily found by taking into account the cubic crystal anisotropy \cite{Bak80}. Thus, there are two possible orientations of the spiral axis, either $\left<100\right>$ or $\left<111\right>$ (Fig.~\ref{fig:helices}).

Upon application of a small magnetic field the helix axis becomes separated from the crystallograpic direction and aligned along the field. A constant average magnetization along the field also appears. This kind of spiral structure is called conical (Fig.~\ref{fig:helices}). The average magnetization increases linearly with the field, and at $H=H_{c2}=D^2M_0/2J$ the cone collapses and the helix goes into a homogeneous state. Note that the wave number of the conical spiral does not depend on the magnetic field.

\subsection{Microscopic details}

\begin{figure}[h]
\includegraphics[width=7cm]{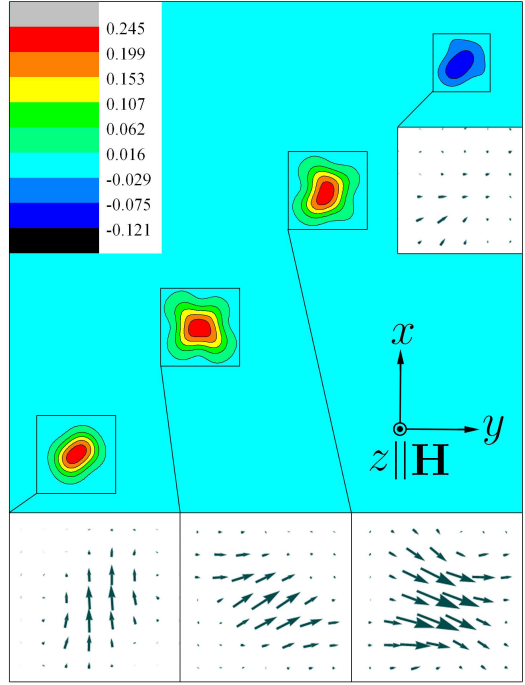}
\caption{\label{fig:quantum} (Color online) The calculated distribution of $m_z$ component of the magnetic moment density in the unit cell of Cu$_2$OSeO$_3$. The magnetic field $\mathbf{H}$ is directed along the $z$ axis. The picture shows the plane $z=0.875$, near which four copper atoms are placed: three Cu-II and one Cu-I (in the top right corner). The insets show distribution of small $m_x$ and $m_y$ components  of the local magnetization, corresponding to spin tilts from the $z$ axis.}
\end{figure}

The phenomenological theory describes magnetic structure as a continuous field of magnetization, $\mathbf{M}(\mathbf{r})$. However, the magnetic moment is not smoothly distributed in a crystalline space, but concentrated mostly near atoms discretely arranged in a lattice. Figure~\ref{fig:quantum} shows a typical  distribution of the magnetic moment density inside the unit cell, calculated with the open-source computer codes for electronic and magnetic structure calculations {\em ``Quantum ESPRESSO''} \cite{QE} (the details of the calculations will be published elsewhere). Thus, the actual magnetic structure has features, which are not described by the continuous model, but can be found in a microscopic approach, such as, for example, the Heisenberg model of chiral ferromagnet with energy \cite{Hopkinson,Shekhtman,Yildirim}
\begin{equation}
\label{eq:Emicro}
E = \sum_{\left\lbrace ij\right\rbrace } \left\lbrace  -J_{ij} (\mathbf{s}_i \cdot \mathbf{s}_j) + \mathbf{D}_{ij} \cdot [\mathbf{s}_i \times \mathbf{s}_j] \right\rbrace - \mathbf{H} \cdot \sum_{i} g_i \mu_B \mathbf{s}_i .
\end{equation}
Here the first sum is taken over bonds $(ij)$ between magnetic atoms, the second one is over the atoms ($i$), $g_i\mu_B\mathbf{s}_i$ is the magnetic moment of the $i$th atom, $\mathbf{s}_i$ is its classical spin ($|\mathbf{s}_i|=1$), $J_{ij}$ and $\mathbf{D}_{ij}$ are parameters characterizing the isotropic exchange and the Dzyaloshinskii--Moriya interaction.

\begin{figure}[h]
\includegraphics[width=7cm]{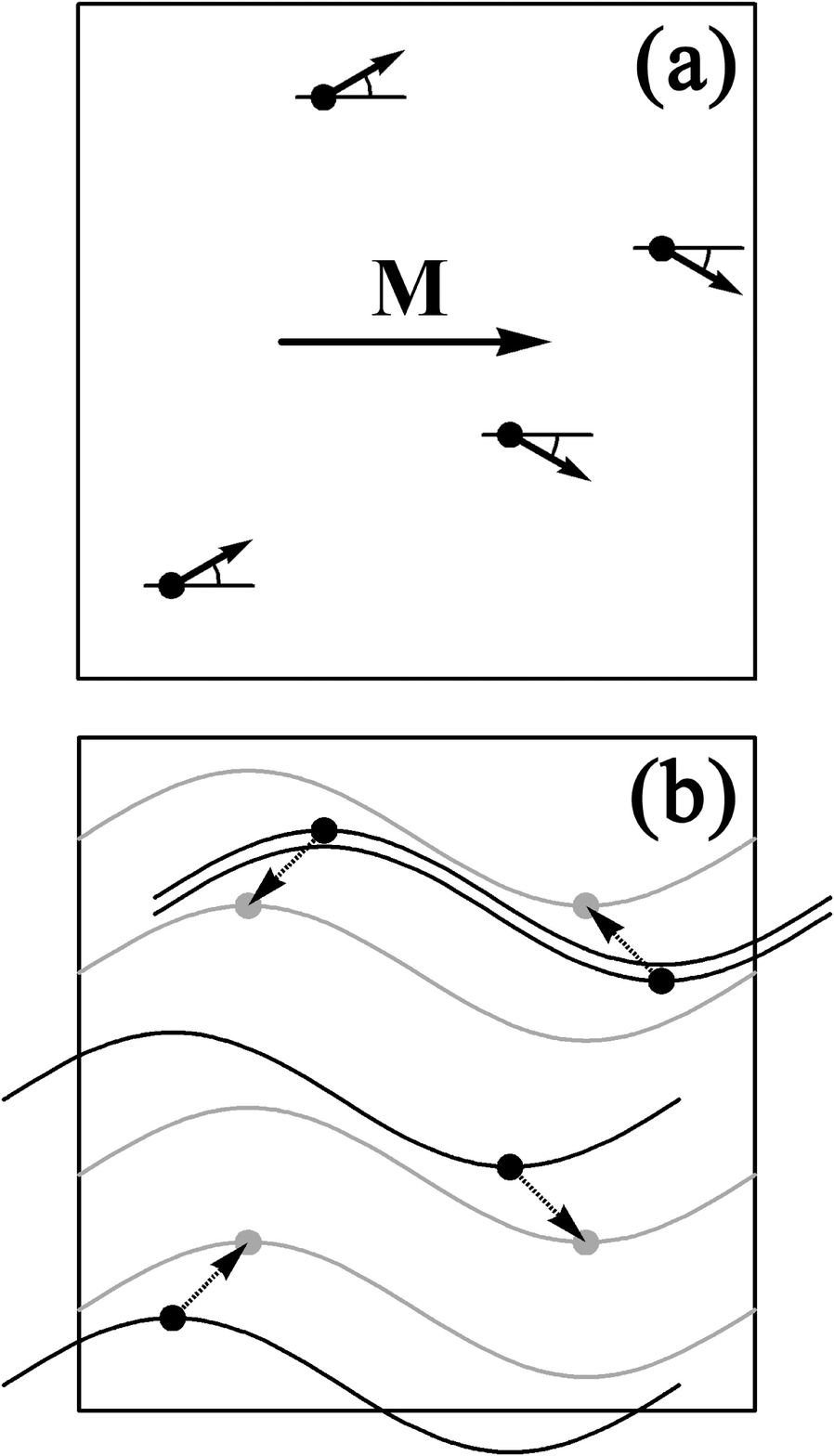}
\caption{\label{fig:canting} Two kinds of cantings. (a) Because of the Dzyaloshinskii--Moriya interaction all the spins are tilted from the magnetization $\mathbf{M}$. (b) Every magnetic position in the unit cell is connected with an individual spin helix; the ``shift'' of the atoms to fictitious positions (shown in gray) makes all the helices co-phased. Shown are the manganese atoms in the MnSi crystal. The directions of cantings and shifts are chosen as in the $B20$ structure; their absolute values, as well as the helix wavenumber are enlarged for visibility.}
\end{figure}

It has been shown in \cite{Dmitrienko2012,Chizhikov2012,Chizhikov2013,Chizhikov2015}, that there is a transition from Eq.~(\ref{eq:Emicro}) to Eq.~(\ref{eq:Ephem}), and phenomenological constants $J$ and $D$ can be calculated as functions of $J_{ij}$ and $\mathbf{D}_{ij}$ at least for smoothly varying magnetic structures. Some features of the spin structure are revealed beyond the continuous model. In particular, it is found that, if the unit cell of a crystal contains several magnetic atoms, then each individual spin is tilted from the average direction of magnetization (Fig.~\ref{fig:canting}a). These tilts, called cantings, in the first approximation, have the form $[\boldsymbol{\rho}_i\times\boldsymbol{\mu}]$, where $\boldsymbol{\mu}=\mathbf{M}/M_0$ is the magnetization direction near the atom, and the ``canting angle'' $\boldsymbol{\rho}_i$ has the following properties: (1) $\boldsymbol{\rho}_i$ possesses the symmetry of the atomic position; for example, for the $B20$ structure, where the atoms lie on the axes 3 of the crystal, the vectors $\boldsymbol{\rho}_i$ are also directed along these axes; (2) canting angles of equivalent atoms in the unit cell are connected to each other by the symmetry transformations of the crystal point group \cite{Chizhikov2015}.

Another kind of cantings is associated with spatial gradients of $\mathbf{M}(\mathbf{r})$. If the magnetization varies smoothly along the crystal, such as forming a magnetic helix, we can naively assume that the classic spins are exactly in accordance with the positions of the atoms along the helix. However, this assumption is not well-founded, which is clear from the fact that the energy (\ref{eq:Emicro}) does not contain an information about the coordinates of the magnetic atoms. Indeed, instead of the single magnetization helix, it is necessary to consider several spin spirals (according to the number of magnetic atoms in the unit cell) with phase shifts between them determined from minimization of Eq.~(\ref{eq:Emicro}) (Fig.~\ref{fig:canting}b). It is known \cite{Chizhikov2015}, that this problem can be solved by assuming that the atoms are placed in some fictitious positions, rather than in their actual positions inside the crystal. In this case, the phase shifts disappear and all the spins lie on a common helix. The fictitious position can be found from the condition
\begin{equation}
\label{eq:exchcoor}
\sum_j J_{ij} c_j \mathbf{r}_{ij} = 0 ,
\end{equation}
where the summation is taken over the magnetic neighbors of the $i$th atom, $c_j=\pm 1$ are some factors, the meaning of which will be explained below, and $\mathbf{r}_{ij}$ is the distance between the fictitious positions of the $i$th and $j$th atoms. The equation (\ref{eq:exchcoor}) is a system of linear equations, with the fictitious coordinates being the unknowns, and all the coefficients and the constant terms being linear combinations of $J_{ij}$. Taking into account the symmetry of the crystal, it is possible to reduce the number of independent coordinates to a minimum (1 for the $B20$ structure, 4 for Cu$_2$OSeO$_3$). Because all the fictitious coordinates are functions of the exchange constants $J_{ij}$, they are also called ``exchange coordinates''.

The absolute value of an atomic magnetic moment in a weak field $H<H_{c2}$ is only slightly depending on the field intensity and the magnetic structure. For example, the magnetic moment of the Cu$^{2+}$ ion in Cu$_2$OSeO$_3$ is about 0.61$\mu_B$ \cite{BosPRB2008}. As a result, in the first approximation, the cantings are small (of the order of $D/J$) corrections to the spins, almost perpendicular to the magnetization direction $\boldsymbol{\mu}$. In addition, the sum of the cantings of all spins in the unit cell, $[\sum_i \boldsymbol{\rho}_i \times \boldsymbol{\mu}]$, is equal to zero by virtue of the cubic symmetry of the crystal. This allows us to consider the cantings as a demonstration of a weak antiferromagnetism on the background of a strong ferromagnetic state. Thus, this case is symmetric to the case of a weak ferromagnet, in which weak cantings against the background of an antiferromagnetic spin order give rise to a small magnetization \cite{Dzyaloshinskii58,Moriya60b}.

Both kinds of cantings discussed above are important for macroscopic properties of the crystal. For example, they are needed for the correct calculation of the phenomenological constants $J$ and $D$ of the model (\ref{eq:Emicro}) \cite{Chizhikov2015}. As will be shown later, the cantings can also give rise to an electric polarization of multiferroics. Recently \cite{DalmasDeReotierPRB2016}, some microscopic details of the spin structure of the helimagnet MnSi were experimentally observed, that could be interpreted as the cantings of the second kind, associated with gradients of the magnetization (Fig.~\ref{fig:canting}b).

Along with the spin cantings, caused by the Dzyaloshinskii--Moriya interaction and the twist of the magnetic structure, there are spin tilts arising from the interaction with a local crystal field. These tilts are of the form $\pm(\boldsymbol{\alpha}_i\cdot\boldsymbol{\mu})[[\boldsymbol{\alpha}_i\times\boldsymbol{\mu}]\times\boldsymbol{\mu}]$, where associated with the atomic positions vectors $\boldsymbol{\alpha}_i$ have the same symmetry properties as $\boldsymbol{\rho}_i$. Note that for the $B20$ structure, where the atoms lie on the axes 3, the vectors $\boldsymbol{\rho}_i$ and $\boldsymbol{\alpha}_i$ are collinear to the axes and, therefore, the cantings owing to the Dzyaloshinskii--Moriya interaction and the local anisotropy should be perpendicular to each other. Below, we will not consider the cantings due to the local crystal field, neglecting their possible participation in the formation of the electric polarization.

\section{Electric polarization in cubic chiral magnets}
\label{sec:epolcub}

Let us consider the cubic crystals of the space group $P2_13$, exemplified by the $B20$ structure magnets (MnSi, MnGe, etc.) and Cu$_2$OSeO$_3$. It is known that due to the chirality of the crystals a magnetic helix can arise, simple or conical (the latter in an external magnetic field), with a helical pitch much greater than the period of the crystal lattice. For instance, in the Cu$_2$OSeO$3$ crystal the helix pitch is about 70 lattice periods. The bonds between the magnetic atoms in the structures do not have symmetry elements, and therefore all $\mathbf{v}$ and $T$ are vectors and pseudotensors of general form.

Because the electric polarization is an additive vector, we can consider only one of the equivalent bonds within the unit cell. In order to calculate the spins, we will replace the discrete magnetic moments by continuous spin fields \cite{Chizhikov2013,Chizhikov2015}. So, the polarization will also be a continuous function of the spatial coordinates.

Suppose that the macroscopic magnetization in atomic position 1 is directed along a unit vector $\boldsymbol{\mu}$. Then the spins of the magnetic atoms in two close positions 1 and 2 can be defined as
\begin{equation}
\label{eq:si}
\mathbf{s}_i \approx c_i (\boldsymbol{\mu} + \mathbf{w}_{i\perp} - w_{i\perp}^2 \boldsymbol{\mu} / 2) , \phantom{x} i = 1, 2 ,
\end{equation}
with $c_i$ being introduced for ferrimagnet Cu$_2$OSeO$_3$ coefficients, considering that the spins in some atomic positions have the opposite direction to the magnetization (Fig.~\ref{fig:cu2oseo3}) \cite{BosPRB2008}, $c_i=-1$ for 4 Cu-I atoms and $c_i=+1$ for 12 Cu-II atoms \cite{Chizhikov2015}; $\mathbf{w}_{i\perp}$ is the perpendicular to $\boldsymbol{\mu}$ spin change of the first order in $D/J$, and the last term in the parentheses represents the reduction of the spin along $\boldsymbol{\mu}$, being of the second order in $D/J$.

The canting of spin 1 is associated with the Dzyaloshinskii--Moriya interaction only,
\begin{equation}
\label{eq:canting1}
\mathbf{w}_{1\perp} = [\boldsymbol{\rho}_1 \times \boldsymbol{\mu}] ,
\end{equation}
whilst spin 2 has an additional tilt due to the global twist of the magnetic structure,
\begin{equation}
\label{eq:canting2}
\mathbf{w}_{2\perp} = [\boldsymbol{\rho}_2 \times \boldsymbol{\mu}] + k (\mathbf{n} \cdot \mathbf{r}_{12}) [\mathbf{n} \times \boldsymbol{\mu}] .
\end{equation}
Here, $k$ and $\mathbf{n}$ are the helix wavenumber and direction, respectively, and $\boldsymbol{\rho}_i$ are the canting angles corresponding to the atomic positions within the unit cell.

Note that vector $\mathbf{r}_{12}$ contains the exchange coordinates of the atoms, which are functions of the isotropic exchange parameters $J_{ij}$, see Eq.~(\ref{eq:exchcoor}). The product $(\mathbf{n}\cdot\boldsymbol{\mu})$ is equal to zero for the simple helices and has a non-zero constant value for each of the conical spirals.

In order to calculate the contribution of the bond to the polarization, it is necessary to average Eq.~(\ref{eq:pbond}) over the operations of the point symmetry group 23. Note that these operations act on vectors $\mathbf{v}$, $\boldsymbol{\rho}_1$, $\boldsymbol{\rho}_2$, $\mathbf{r}_{12}$ and pseudotensor $T$, whereas $\mathbf{n}$ and $\boldsymbol{\mu}$ assumed to be invariant for the given point. We also can use the weakness of the Dzyaloshinskii--Moriya interaction ($D$), as compared with the isotropic exchange ($J$):
\begin{equation}
\label{eq:small}
\rho \sim k \sim T/v \sim D/J \ll 1 .
\end{equation}
Upon averaging, the terms of the zero and first orders in $D/J$ disappear. We confine ourselves to second-order terms. Then averaging of Eq.~(\ref{eq:pbond}) gives
\begin{equation}
\label{eq:pmean}
\begin{array}{l}
\left<\mathbf{p}\right> = \mathbf{p}_1 + \mathbf{p}_2 + \mathbf{p}_3 + \mathbf{p}_4 , \\
\mathbf{p}_1 = a (\mu_y \mu_z, \mu_z \mu_x, \mu_x \mu_y) , \\
\mathbf{p}_2 = b k \left\{ (n_y n_z, n_z n_x, n_x n_y) - (\mathbf{n} \cdot \boldsymbol{\mu}) (n_y \mu_z, n_z \mu_x, n_x \mu_y) \right\} , \\
\mathbf{p}_3 = c k \left\{ (n_y n_z, n_z n_x, n_x n_y) - (\mathbf{n} \cdot \boldsymbol{\mu}) (\mu_y n_z, \mu_z n_x, \mu_x n_y) \right\} , \\
\mathbf{p}_4 = d k^2 \left\{ 1 - (\mathbf{n} \cdot \boldsymbol{\mu})^2 \right\} (n_y n_z, n_z n_x, n_x n_y) .
\end{array}
\end{equation}
The coefficients $a$, $b$, $c$, and $d$ are calculated by the formulas
\begin{equation}
\label{eq:abcd}
\begin{array}{l}
a = c_1 c_2 \left\{ [{\cal R} + {\cal L}](\mathbf{v} \otimes \Delta\boldsymbol{\rho} \otimes \Delta\boldsymbol{\rho}) / 6 - [{\cal R} + {\cal L}](T \otimes \Delta\boldsymbol{\rho}) / 3 \right\} , \\
b = c_1 c_2 \left\{ -{\cal R}(\mathbf{v} \otimes \mathbf{r}_{12} \otimes \Delta\boldsymbol{\rho}) + {\cal L}(T \otimes \mathbf{r}_{12}) \right\} / 3 , \\
c = c_1 c_2 \left\{ -{\cal L}(\mathbf{v} \otimes \mathbf{r}_{12} \otimes \Delta\boldsymbol{\rho}) + {\cal R}(T \otimes \mathbf{r}_{12}) \right\} / 3 , \\
d =  - c_1 c_2 [{\cal R} + {\cal L}](\mathbf{v} \otimes \mathbf{r}_{12} \otimes \mathbf{r}_{12}) / 6 ,
\end{array}
\end{equation}
with $\Delta\boldsymbol{\rho} = \boldsymbol{\rho}_2 - \boldsymbol{\rho}_1$, ${\cal R}(A) = A_{xyz} + A_{yzx} + A_{zxy}$, ${\cal L}(A) = A_{xzy} + A_{yxz} + A_{zyx}$, and $[{\cal R} + {\cal L}](A) = {\cal R}(A) + {\cal L}(A)$. Note that if, as in this case, 3rd rank tensor $A$ has the form $M\otimes\mathbf{a}$, or, in particular, $\mathbf{a}\otimes\mathbf{b}\otimes\mathbf{c}$, where $M$ is a 2nd rank tensor, then ${\cal R}(A)$ and ${\cal L}(A)$ are invariants of the symmetry operations of the point group 23 and, thus, do not depend on our choice of an equivalent bond.

All the terms in Eq.~(\ref{eq:pmean}) are of the same order in small parameter $D/J$, but it is also possible that another small parameter exists, arising from symmetry considerations. There is an obvious symmetry between terms $\mathbf{p}_2$ and $\mathbf{p}_3$ and, in particular, between coefficients $b$ and $c$. Indeed, ${\cal R}(A)$ and ${\cal L}(A)$ can be related together by some symmetry operation outside of point group 23, which changes the parity of permutation $xyz$. So, the rotation by angle $\pi$ about the axis [110] transforms ${\cal R}$ into $-{\cal L}$, and ${\cal L}$ into $-{\cal R}$. This rotation enhances the symmetry to the point group 432, also without an inversion center. Then, some of the coefficients degenerate: $a=b+c=d=0$, and the average polarization takes an ``isotropic'' form (i.e., invariant with respect to the rotation of the Cartesian axes),
\begin{equation}
\label{eq:piso}
\left<\mathbf{p}\right> \sim k (\mathbf{n} \cdot \boldsymbol{\mu}) [\mathbf{n} \times \boldsymbol{\mu}] .
\end{equation}
The smallness of coefficients $a$, $b+c$, $d$ depends, therefore, on how close the symmetry of the crystal is to the point group 432. It has been shown in \cite{Chizhikov2015} that the cubic crystal Cu$_2$OSeO$_3$, first described in \cite{Effenberger1986}, can be seen as deformed structure of a crystal with the space group $P4_132$. Because this group containes right-handed axes $4_1$, it is natural to call the crystal described in \cite{Effenberger1986} a right one, and its mirror-image enantiomer a left one. The left crystal Cu$_2$OSeO$_3$ also possesses the spatial symmetry $P2_13$, but its structure is approximated by the high-symmetry group $P4_332$ with left-handed axes $4_3$. We can assume that for Cu$_2$OSeO$_3$ coefficients $a$, $b+c$, $d$ of the anisotropic terms can be small. Nevertheless, it is possible that there are crystals of the group $P2_13$, whose structures are not approximated by high-symmetry groups, and these coefficients are large.

\begin{figure}[h]
\includegraphics[width=7cm]{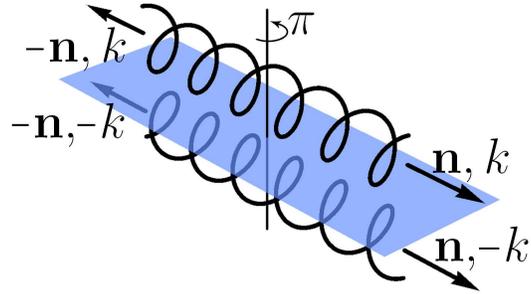}
\caption{\label{fig:kn} (Color online) Simple helix described by a wavenumber $k$ and a axis direction $\mathbf{n}\equiv -\mathbf{n}$. The rotation by angle $\pi$ around the perpendicular axis does not change the helix, but replace $\mathbf{n}$ by $-\mathbf{n}$. The reflection in a plane parallel to the helix changes the sign of chirality, $k\rightarrow -k$, transforming the right spiral to the left one, but keeps vector $\mathbf{n}$ unchanged.}
\end{figure}

It is useful to explore the symmetry properties of Eqs.~(\ref{eq:pmean}) and (\ref{eq:abcd}) under inversion, changing the chirality of the crystal. The quantities determining the coefficients (\ref{eq:abcd}) possess the following properties: $\mathbf{v}$ and $\mathbf{r}_{12}$ are vectors, $T$ is a pseudotensor, the canting angles $\boldsymbol{\rho}_i$ are axial vectors (pseudovectors). It is easy to find that $b$ and $c$ are scalars, whereas $a$ and $d$ are pseudoscalars. The magnetization direction $\boldsymbol{\mu}$ changes its sign under inversion, but this does not affect the electric polarization, because it contains $\boldsymbol{\mu}$ in quadratic combinations due to the symmetry under time reversal. Direction $\mathbf{n}$ of the helix axis also appeares in the formulas only in even powers, albeit for a different reason. The simple helix is described by a wavenumber $k$ and a direction $\mathbf{n}$, besides $k$ can take both positive and negative values, thereby determining the sign of chirality. In turn, $\mathbf{n}$ has the same properties as the director of a cholesteric liquid crystal: the replacement of $\mathbf{n}$ with $-\mathbf{n}$ does not change the spiral (Fig.~\ref{fig:kn}). So, the invariance of coefficients $b$ and $c$ is compensated by the change of the sign of $k$, and, as expected, the electric polarization changes the sign under inversion. All the above can be summarized as follows:
\begin{equation}
\label{eq:inversion}
\begin{array}{c}
I: \\
\mathbf{v} \rightarrow -\mathbf{v}, \phantom{x} \mathbf{r}_{12} \rightarrow -\mathbf{r}_{12}, \phantom{x} T \rightarrow -T, \phantom{x} \boldsymbol{\rho}_i \rightarrow \boldsymbol{\rho}_i, \\
a \rightarrow -a, \phantom{x} b \rightarrow b, \phantom{x} c \rightarrow c, \phantom{x} d \rightarrow -d, \\
\boldsymbol{\mu} \rightarrow -\boldsymbol{\mu}, \phantom{x} \mathbf{n} \rightarrow \pm\mathbf{n}, \phantom{x} k \rightarrow -k, \phantom{x} \mathbf{p} \rightarrow -\mathbf{p} .
\end{array}
\end{equation}

All the arguments given above can be repeated for the total electric polarization $\mathbf{P}=\sum f_i \left<\mathbf{p}_i\right>$, where the sum is taken over all crystallographically nonequivalent bonds between magnetic atoms, $f_i$ is the multiplicity of the $i$th bond within the unit cell.

\section{Electric polarization averaged over crystal}
\label{sec:averagepol}

Described by Eq.~(\ref{eq:piso}) part of the electric polarization vanishes both for the cases of a simple helix ($\mathbf{n}\perp\boldsymbol{\mu}$) and a homogeneous magnetization ($\mathbf{n}\parallel\boldsymbol{\mu}$), and has a maximum value in the external magnetic field with magnitude of $H_{c2}/\sqrt{2}$, when the angle between vectors $\mathbf{n}$ and $\boldsymbol{\mu}$ is equal to $\pi/4$. In addition, when moving along the helix, the polarization vector described by Eq.~(\ref{eq:piso}) rotates in the plane perpendicular to $\mathbf{n}$, therefore its average value is zero. However, in the case of the space group $P2_13$, there exists an electric polarization associated with the crystal anisotropy, and its average value over the crystal can differ from zero.

Averaging of Eq.~(\ref{eq:pmean}) over the crystal gives
\begin{equation}
\label{eq:Pcryst}
\left<\mathbf{P}\right> = \{[D k^2 + B k + C k - A/2] - [D k^2 + B k + C k - 3A/2] (\mathbf{n} \cdot \boldsymbol{\mu})^2\} (n_y n_z, n_z n_x, n_x n_y) ,
\end{equation}

where constants $A$, $B$, $C$, $D$ arise from $a$, $b$, $c$, $d$ after summing over all the bonds in the unit cell. Recall that Eq.~(\ref{eq:pmean}), and hence Eq.~(\ref{eq:Pcryst}), is derived for a conical spiral in an external magnetic field, particular cases of which are the simple helix at $H=0$ and the homogeneous state at $H>H_{c2}$. The product $(\mathbf{n}\cdot\boldsymbol{\mu})$ changes linearly from 0 to 1 with increase of $H$ from 0 to $H_{c2}$, and thus, the electric polarization magnitude $|\left<\mathbf{P}\right>|$ is a quadratic function of the field. Measurements of $|\left<\mathbf{P}\right>|$ for different magnetic field values allow us to distinguish $A$ and $Dk^2+Bk+Ck$. In principle, the separation of coefficient $D$ and $B+C$ is also possible, measuring the electric polarization for spirals with different wavenumbers. The difficulty here is that the equilibrium value of $k$ is determined by the ratio of the parameters of the isotropic exchange and the Dzyaloshinskii--Moriya interaction. The spirals with different $k$ can be obtained, for example, by fixing the magnetization direction at the boundaries of a thin crystalline sample. Finally, Eq.~(\ref{eq:Pcryst}) does not imply a possibility of separation of coefficients $B$ and $C$; this will require a more subtle experiment to measure the local electric polarization $\mathbf{P}$.

\subsection{Average electric polarization of the A-phase}

\begin{figure}[h]
\includegraphics[width=7cm]{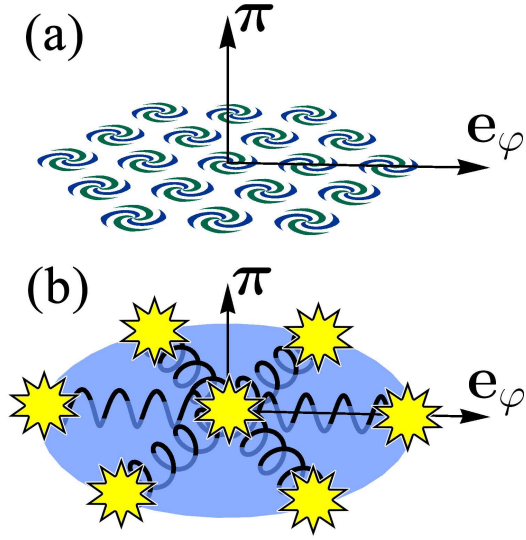}
\caption{\label{fig:Aphase}  (Color online) (a) The A-phase is a triangular lattice of 2D skyrmion vortices aligned along vector $\boldsymbol{\pi}$, coinciding with the direction of an external magnetic field. The orientation of the lattice in the plane perpendicular to $\boldsymbol{\pi}$ is defined by a vector $\mathbf{e}_\varphi$. (b) The observed diffraction pattern consists of a central peak (scattering at zero angle) and six symmetrically arranged peaks. It allows us to describe the A-phase as a superposition of three spirals with the wavenumber close to that of the helical structure.}
\end{figure}

The A-phase possesses a hexagonal symmetry and more complex structure compared to the helices \cite{Bogdanov89,Rossler06,Ambrose2013}. There is an energy gain in the A-phase due to the double twist of the magnetization direction $\boldsymbol{\mu}$. Whilst the spatial orientation of a helix is defined only by vector $\mathbf{n}$, in order do describe the A-phase, two perpendicular vectors are needed: $\boldsymbol{\pi}$, directed along an applied magnetic field and perpendicular to the plane of vortices, and $\mathbf{e}_\varphi$, specifying an azimuthal rotation angle of the triangular lattice of vortices (Fig.~\ref{fig:Aphase}a). As in the case of a conical spiral, the average magnetic moment is different from zero and directed along $\boldsymbol{\pi}$. The diffraction pattern contains six symmetrically spaced peaks with a wavenumber close to that of the helical structure (Fig.~\ref{fig:Aphase}b). We can assume that, in the simplest approximation, the A-phase is a superposition of a homogeneous magnetization field with $\boldsymbol{\mu}\parallel\boldsymbol{\pi}$ and three perpendicular to $\boldsymbol{\pi}$ helices with the angles of $2\pi/3$ between them. Assuming the additivity of the electric polarization, it is easy to find that the average polarization does not depend on the azimuthal angle and is determined only by $\boldsymbol{\pi}$, $\left<\mathbf{P}\right>\sim(\pi_y\pi_z,\pi_z\pi_x,\pi_x\pi_y)$.

\subsection{The angular dependence of the electric polarization}

\begin{figure}[h]
\includegraphics[width=7cm]{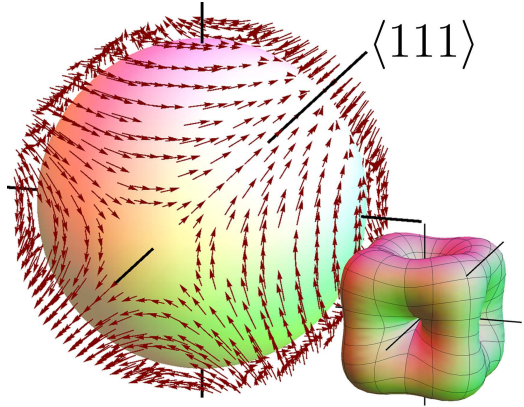}
\caption{\label{fig:absp} (Color online) The dependence of the average electric polarization and its absolute value (in the corner) on the orientation of the magnetic structure with respect to the crystallographic axes in a chiral cubic multiferroic with the spase group $P2_13$. The polarization is maximal for the crystallographic directions $\left<111\right>$ and zero for $\left<100\right>$.}
\end{figure}

Summarizing the results of this section, the average electric polarization for different magnetic structures observed in the cubic multiferroics with the space group $P2_13$ can be written in general form as
\begin{equation}
\label{eq:Pcommon}
\left<\mathbf{P}\right> \sim (e_y e_z, e_z e_x, e_x e_y) ,
\end{equation}
where unit vector $\mathbf{e}$ can take the values $\boldsymbol{\mu}$, $\mathbf{n}$, and $\boldsymbol{\pi}$, respectively, for a homogeneous magnetization field, a single spiral, and the A-phase. The coordinates of $\mathbf{e}$ are defined in the Cartesian system associated with the cubic crystal lattice. It can be seen that the angular dependence of the average polarization on the direction of $\mathbf{e}$ is strongly anisotropic. Thus, the polarization magnitude
\begin{equation}
\label{eq:absPcommon}
|\left<\mathbf{P}\right>| \sim \sqrt{1 - e^4_x - e^4_y - e^4_z}
\end{equation}
varies from zero for $\mathbf{e} \sim \left<100\right>$ to its maximum value for $\mathbf{e} \sim \left<111\right>$ (Fig.~\ref{fig:absp}). In principle, the average electric polarization $\left<\mathbf{P}\right>$ can be at an arbitrary angle to $\mathbf{e}$. For example, the polarization is parallel to $\mathbf{e}$ for directions $\left<111\right>$, and perpendicular to it for directions $\left<110\right>$.

In the case of a homogeneous magnetization field $\boldsymbol{\mu}$, Eq.~(\ref{eq:Pcommon}) has been derived in \cite{Mochizuki2015} using a single-spin model and confirmed experimentally for Cu$_2$OSeO$_3$ crystal \cite{SekiSci2012,SekiPRB2012-2}.

\section{The phenomenological expression for electric polarization}
\label{sec:phenom}

The equation (\ref{eq:pmean}) for the electric polarization is derived for the case of a spiral magnetic structure and contains the wavenumber $k$ and the helix axis direction $\mathbf{n}$. The helical twist, however, is not the only possible type of spin ordering, and, in particular,  the A-phase in MnSi and Cu$_2$OSeO$_3$ crystals possesses a complex magnetic structure with a double twist of the magnetization \cite{Bogdanov89,Rossler06,Ambrose2013}. In general, an arbitrary field $\mathbf{M}(\mathbf{r})$ should be considered with a possible restriction concerning the constancy of its magnitude, $\mathbf{M}(\mathbf{r})=M_0\boldsymbol{\mu}(\mathbf{r})$. In this case it is necessary to replace Eq.~(\ref{eq:pmean}) by expressions for the electric polarization, depending only on $\boldsymbol{\mu}$ and its spatial derivatives. These expressions can serve as a basis for a continual phenomenological theory, which describes the appearance of ferroelectricity as a result of a magnetic ordering in the crystal.

In the case of a conical spiral, the derivatives of the magnetization direction $\boldsymbol{\mu}$ are of the form
\begin{equation}
\label{eq:dmudr}
\frac{\partial \boldsymbol{\mu}}{\partial r_\alpha} = k n_\alpha [\mathbf{n} \times \boldsymbol{\mu}] , \phantom{x} \alpha = x, y, z .
\end{equation}
Using Eq.~(\ref{eq:dmudr}), it is easy to find a generalization of Eq.~(\ref{eq:pmean}) for an arbitrary field $\boldsymbol{\mu}(\mathbf{r})$:
\begin{equation}
\label{eq:Pmean2}
\begin{array}{l}
\mathbf{P} = \mathbf{P}_{iso} + \mathbf{P}_A + \mathbf{P}_{BC} + \mathbf{P}_D , \\
\mathbf{P}_{iso} = \frac{B - C}2 \{ \boldsymbol{\mu} (\boldsymbol{\nabla} \cdot \boldsymbol{\mu}) - (\boldsymbol{\mu} \cdot \boldsymbol{\nabla}) \boldsymbol{\mu} \} , \\
\mathbf{P}_A = A (\mu_y \mu_z, \mu_z \mu_x, \mu_x \mu_y) , \\
\mathbf{P}_{BC} = \frac{B + C}2 \left( \begin{array}{ll}
\mu_x \frac{\partial \mu_y}{\partial y} - \mu_y \frac{\partial \mu_x}{\partial y} - \mu_x \frac{\partial \mu_z}{\partial z} + \mu_z \frac{\partial \mu_x}{\partial z} \\
\mu_y \frac{\partial \mu_z}{\partial z} - \mu_z \frac{\partial \mu_y}{\partial z} - \mu_y \frac{\partial \mu_x}{\partial x} + \mu_x \frac{\partial \mu_y}{\partial x} \\
\mu_z \frac{\partial \mu_x}{\partial x} - \mu_x \frac{\partial \mu_z}{\partial x} - \mu_z \frac{\partial \mu_y}{\partial y} + \mu_y \frac{\partial \mu_z}{\partial y}
\end{array} \right) , \\
\mathbf{P}_D = D \left( \frac{\partial \boldsymbol{\mu}}{\partial y} \cdot \frac{\partial \boldsymbol{\mu}}{\partial z} , \frac{\partial \boldsymbol{\mu}}{\partial z} \cdot \frac{\partial \boldsymbol{\mu}}{\partial x} , \frac{\partial \boldsymbol{\mu}}{\partial x} \cdot \frac{\partial \boldsymbol{\mu}}{\partial y} \right) .
\end{array}
\end{equation}
Here $\mathbf{P}_{iso}$ is corresponding to Eq.~(\ref{eq:piso}) ``isotropic'' contribution, which does not depend on the orientation of the magnetic structure with respect to the crystallographic axes. On the other hand, terms $\mathbf{P}_A$, $\mathbf{P}_{BC}$, and $\mathbf{P}_D$  are associated with the cubic anisotropy of the considered crystals (vector $\mathbf{P}_{BC}$ written in a column for convenience). All contributions are of the same order in $D/J$, however, as has been discussed above, the anisotropic terms may have an additional degree of smallness for symmetry reasons. Thus, $\mathbf{P}_A$, $\mathbf{P}_{BC}$, and $\mathbf{P}_D$ are nonzero for the crystals of the point group 23, but disappear when the symmetry is enhanced to 432.

In order to obtain Eq.~(\ref{eq:Pmean2}) within the framework of the phenomenological theory, the free energy expression should contain the term of the form
\begin{equation}
\label{eq:F}
F_{\mathbf{P}} \sim \mathbf{P}^2/2 - \mathbf{P} \cdot (\mathbf{P}_{iso} + \mathbf{P}_A + \mathbf{P}_{BC} + \mathbf{P}_D) .
\end{equation}

\section{Discussion}
\label{sec:discussion}

In \cite{Mostovoy2006,PyatakovUFN2015} an expression for the electric polarization is found, which coincides with term $\mathbf{P}_{iso}$ in Eq.~(\ref{eq:Pmean2}). It was shown above that $\mathbf{P}_{iso}$ is the main contribution to the local polarization, but, under the condition that $\boldsymbol{\nabla}\cdot\mathbf{M}=0$, performed in some important cases, such as simple and conical helical magnetic structures, $\mathbf{P}_{iso}$ reduces to a surface term and its impact to the average electric polarization becomes negligible. Here we show that, along with the polarization $\mathbf{P}_{iso}$ for an isotropic magnetic medium, there exist terms associated with the crystal anisotropy, cubic in our case. These terms ($\mathbf{P}_A$, $\mathbf{P}_{BC}$, $\mathbf{P}_D$) cause a nonzero electric polarization when averaged over the crystal, which allows to control the magnetic structure of multiferroic Cu$_2$OSeO$_3$ by the electric field.

An alternative microscopic approach, which takes into account the anisotropy of cubic crystal, is developed in \cite{Mochizuki2015}. In particular, the expression (\ref{eq:Pcommon}) is found, connecting the electric polarization with the magnetization at a point. However, in order to derive Eq.~(\ref{eq:Pcommon}) without taking into account the spin cantings, a single-spin model is used, with the local polarization being calculated by summing the terms of form $\mathbf{r}_{ij}(\mathbf{s}_i\mathbf{r}_{ij})^2$, where $\mathbf{s}_i$ is the spin of the $i$th ion Cu$^{2+}$ and $\mathbf{r}_{ij}$ is the distance from the copper atom to the neighboring $j$th ion O$^{2-}$ \cite{SekiPRB2012-2,Mochizuki2015,Ruff2015}. A more general single-spin model is used in \cite{GongXiang2012}. As a result, the local polarization has no the gradient part, including the largest contribution $\mathbf{P}_{iso}$. In the present paper, we use a model that takes into account the spin noncollinearity due to the cantings in order to derive terms $\mathbf{P}_{iso}$, $\mathbf{P}_{BC}$, $\mathbf{P}_D$, containing spatial derivatives of the magnetization. The choice between the single-spin and two-spin models can be made experimentally. This requires measuring the average electric polarization of the conical magnetic helix occuring in a bulk sample in an external magnetic field with intensity varying from zero to $H_{c2}$. To compare with theory, Eq.~(\ref{eq:Pcryst}) can be used, with the single-spin model corresponding to the additional condition $B=C=D=0$. The experiment will separate the values of $A$ and $Dk^2+Bk+Ck$, and, if the latter is not zero, the two-spin model of the appearance of the electric polarization is needed.

The appearance of ferroelectricity in the chiral type-II multiferroics can be considered within the framework of a phenomenological approach \cite{Marchenko2014,PikinJETPLett2012,PikinPRB2012,PikinJETP2013,Lyubutin2013,PikinJETPLett2014}. The method is especially effective to describe the phase transition from the paramagnetic phase to a low-temperature multiferroic state. In this case, the expression for the free energy contains terms permitted by the crystal point symmetry, which connect the electric polarization $\mathbf{P}$ with the magnetization $\mathbf{M}$ and its spatial derivatives. Besides, as a rule, restrictions are imposed on the power and the number of derivatives, reflecting the characteristic smallness of spatial gradients. For the chiral magnets discussed in the paper it is determined by the smallness of the Dzyaloshinskii--Moriya interaction as compared with the isotropic exchange, $D/J\ll 1$. The result obtained from our microscopic consideration shows that terms with different numbers of derivatives may be of the same order in $D/J$. Thus, Eq.~(\ref{eq:Pmean2}) contains close in size terms $\mathbf{P}_A$, $\mathbf{P}_{BC}$, and $\mathbf{P}_D$ with 0, 1 and 2 derivatives, respectively. These terms should be taken into account in the phenomenological description, e.g., by including the term (\ref{eq:F}) into the free energy.

\section*{Acknowledgements}

We are grateful to S.~A. Pikin for useful discussions. The reported study was funded by RFBR according to the research project No. 14-02-00268~a.


\newpage
\section*{Figures}

\end{document}